\documentclass[conference]{IEEEtran}
\usepackage[noadjust]{cite}
\usepackage{graphicx,color}
\usepackage[dvipsnames]{xcolor}
\usepackage{amsmath,amsbsy,amsfonts,amssymb,amsthm}
\usepackage{mathrsfs,bm,bbm}
\usepackage{mathtools}
\usepackage{lipsum}
\usepackage{flushend}

\mathtoolsset{showonlyrefs}
\ifCLASSOPTIONcompsoc
\usepackage[caption=false,font=normalsize,labelfont=sf,textfont=sf]{subfig}
\else
\usepackage[caption=false,font=footnotesize]{subfig}
\fi

\IEEEoverridecommandlockouts 

\newcommand{\figr}{Fig.~}
\newcommand{\secr}{Sec.~}

\usepackage{tikz}
\usetikzlibrary{shapes}

\usepackage{acro}
\DeclareAcronym{AWGN}{short = AWGN ,long = additive white gaussian noise}
\DeclareAcronym{AoI}{short = AoI ,long = age of information}
\DeclareAcronym{AoII}{short = AoII ,long = age of incorrect information}

\DeclareAcronym{CDF}{short = CDF ,long = cumulative distribution function}
\DeclareAcronym{CRA}{short = CRA ,long = contention resolution ALOHA}
\DeclareAcronym{CRDSA}{short = CRDSA ,long = contention resolution diversity slotted ALOHA}
\DeclareAcronym{CSA}{short = CSA ,long = coded slotted ALOHA}
\DeclareAcronym{C-RAN}{short = C-RAN ,long = cloud radio access network}
\DeclareAcronym{DAMA}{short = DAMA ,long = demand assigned multiple access}
\DeclareAcronym{DSA}{short = DSA ,long = diversity slotted ALOHA}
\DeclareAcronym{eMBB}{short = eMBB ,long = enhanced mobile broadband}
\DeclareAcronym{FEC}{short = FEC ,long = forward error correction}
\DeclareAcronym{GEO}{short = GEO ,long = geostationary orbit}
\DeclareAcronym{GF}{short = GF ,long = generating function}
\DeclareAcronym{IC}{short = IC ,long = interference cancellation}
\DeclareAcronym{IoT}{short = IoT ,long = Internet of things}
\DeclareAcronym{IRSA}{short = IRSA ,long = irregular repetition slotted ALOHA}
\DeclareAcronym{KPI}{short = KPI ,long = key performance indicator}
\DeclareAcronym{LEO}{short = LEO ,long = low Earth orbit}
\DeclareAcronym{MAC}{short = MAC ,long = medium access}
\DeclareAcronym{mMTC}{short = mMTC ,long = massive machine-type communications}
\DeclareAcronym{MC}{short = MC ,long = Markov chain}
\DeclareAcronym{PDF}{short = PDF ,long = probability density function}
\DeclareAcronym{PER}{short = PER ,long = packet error rate}
\DeclareAcronym{PLR}{short = PLR ,long = packet loss rate}
\DeclareAcronym{PMF}{short = PMF ,long = probability mass function}
\DeclareAcronym{RA}{short = RA ,long = random access}
\DeclareAcronym{rv}{short = r.v. ,long = random variable}
\DeclareAcronym{SA}{short = SA , long = slotted ALOHA}
\DeclareAcronym{SIC}{short = SIC ,long = successive interference cancellation}
\DeclareAcronym{SNR}{short = SNR ,long = signal-to-noise ratio}
\DeclareAcronym{SFG}{short = SFG ,long = signal flow graph}
\DeclareAcronym{TDM}{short = TDM ,long = time division multiplexing}
\begin{document}

\title{\huge On the Role of Early-Termination for Age of Information in Tree-Based Random Access Protocols}

\author{\IEEEauthorblockN{Andrea Munari\IEEEauthorrefmark{2}, \v{C}edomir Stefanovi\'c\IEEEauthorrefmark{3}}
\IEEEauthorblockA{\IEEEauthorrefmark{2} Institute of Communications and Navigation, 
German Aerospace Center (DLR), We\ss ling, Germany\\
\IEEEauthorrefmark{3} Department of Electronic Systems, Aalborg University, Denmark.\\
email: andrea.munari@dlr.de, cs@es.aau.dk}
}

\maketitle
\thispagestyle{empty}
\pagestyle{empty}

\begin{abstract}
    Age of Information (AoI) has emerged as a key metric for assessing data freshness in IoT applications, where a large number of devices report time-stamped updates to a monitor. Such systems often rely on random access protocols based on variations of ALOHA at the link layer, where collision resolution algorithms play a fundamental role to enable reliable delivery of packets. In this context, we provide the first analytical characterization of average AoI for the classical Capetanakis tree-based algorithm with gated access \cite{Capetanakis79}  under exogenous traffic, capturing the protocol's dynamics, driven by sporadic packet generation and variable collision resolution times. We also explore a variant with early termination, where contention is truncated after a maximum number of slots even if not all users are resolved. The approach introduces a fundamental trade-off between reliability and timeliness, allowing stale packets to be dropped to improve freshness. %By highlighting some non-trivial take-aways, our study offers practical insights for AoI-aware protocol design.
\end{abstract}

\newtheorem{prop}{Proposition}
\newtheorem{lemma}{Lemma}
\newtheorem{remark}{Remark}

% maths and probability
\newcommand{\pr}{\ensuremath{\mathsf P}}
\newcommand{\expOp}{\ensuremath{\mathbb E}}
\newcommand{\de}{\mathrm{d}}
\newcommand{\given}{\, | \,}
\newcommand{\givenS}{\ensuremath{\vert}}
\newcommand{\norm}[2]{\left \lVert #1 \right \rVert_{#2}}

% Markov chains
\newcommand{\pTrans}{\ensuremath{q}}

\newcommand{\Mc}{\ensuremath{X}}
\newcommand{\Mcn}{\ensuremath{X_n}}
\newcommand{\Rc}{\ensuremath{Y}}
\newcommand{\rc}{\ensuremath{y}}
\newcommand{\Rcn}{\ensuremath{\Rc_n}}
\newcommand{\rcn}{\ensuremath{\rc_n}}
\newcommand{\Distn}{\ensuremath{D_n}}
\newcommand{\distn}{\ensuremath{d_n}}
\newcommand{\Est}{\ensuremath{\hat{X}}}
\newcommand{\Estn}{\ensuremath{\hat{X}_n}}
\newcommand{\mc}{\ensuremath{x}}
\newcommand{\mcn}{\ensuremath{\mc_n}}
\newcommand{\est}{\ensuremath{\hat{x}}}
\newcommand{\estn}{\ensuremath{\hat{x}_n}}

\newcommand{\Reset}{\ensuremath{Z}}
\newcommand{\Refresh}{\ensuremath{Y}}
\newcommand{\cri}{\ensuremath{\mathsf C}}

\newcommand{\Lmax}{\ensuremath{L_{\mathsf{m}}}}
\newcommand{\pGen}{\ensuremath{\rho}}
\newcommand{\pGenCRI}{\ensuremath{\Gamma}}
\newcommand{\LCRI}{\ensuremath{L^\prime}}
\newcommand{\lCRI}{\ensuremath{\ell^\prime}}

% channel access
\newcommand{\pTx}{\ensuremath{\zeta}}
\newcommand{\ps}{\ensuremath{\mathsf{p_s}}}
\newcommand{\peras}{\ensuremath{\varepsilon}}
\newcommand{\nodes}{\ensuremath{\mathsf U}}

% basic metrics
\newcommand{\tru}{\ensuremath{\mathsf S}}
\newcommand{\load}{\ensuremath{\mathsf G}}
\newcommand{\Age}{\ensuremath{\Delta}}
\newcommand{\Agen}{\ensuremath{\Delta_n}}
\newcommand{\agen}{\ensuremath{\delta_n}}
\newcommand{\ent}{\ensuremath{H}}
\newcommand{\condent}{\ensuremath{\mathsf h}}

% AoI related metrics
\newcommand{\overbar}[1]{\mkern 1.5mu\overline{\mkern-2mu#1\mkern-7mu}\mkern .5mu}
\newcommand{\overbara}[1]{\mkern 1.5mu\overline{\mkern0.1mu#1\mkern-0.1mu}\mkern 1.5mu}

\section{Introduction} \label{sec:intro}

Age of information (AoI) has recently emerged as an insightful performance indicator for many communications networks \cite{Yates20_Survey}. Defined as the time elapsed since the generation of the last received message, the metric captures in a simple way how fresh the knowledge available at the receiver is on the status of the source. AoI is especially relevant in Internet of things (IoT) systems, where devices sample a physical process of interest and send time-stamped updates to a central node for monitoring/actuation, and where timeliness is paramount. 

IoT networks are often characterized by a large number of reporting terminals, which generate messages in a sporadic and unpredictable fashion. To accomodate such traffic, many commercial solutions, e.g., LoRa, Sigfox or, partly, NB-IoT, implement random access protocols based on variations of ALOHA \cite{Abramson77:PacketBroadcasting} at the link layer. In parallel, research has extensively focused on devising solutions that ease the well-known issue of collisions in grant-free channels. On the one hand, a vast wave of new schemes, often referred to as modern- or unsourced-random access \cite{Liva24_Proceedings}, constructively embrace interference and decode users resorting to advanced physical layer techniques. On the oher hand, a number of solutions have been proposed to efficiently resolve users via retransmissions of collided packets. Among these, a smart approach is offered by tree-based algorithms. Remaining true to the random-access paradigm, these schemes rely on feedback and provide simple, recursive, procedures that users can implement after a collision, until each of the involved senders manages to deliver its message. Originally introduced by Capetanakis \cite{Capetanakis79}, the idea has been developed thoroughly, see, e.g., \cite{Massey81,Mathys85,Giannakis07,Stefanovic24} and references therein, with particular attention to improving throughput.

Notably, while the performance of random access solutions is well-understood for traditional metrics, more limited results are available in terms of AoI. Initial works \cite{Yates17:AoI_SA,Yates20_ISIT} explored the behavior of ALOHA and some age-specific optimizations, e.g., \cite{Uysal21_AlohaThresh,Bidokhti22_TIT} among others, whereas some recent works tackled modern random access schemes, e.g., \cite{Munari21_TCOM_AoI,Munari23_TCOM}. On the other hand, a characterization of tree-based algorithms remains largely unexplored, with early results \cite{Pan22_TVT} considering the generate-at-will case where nodes have new messages to transmit all the time. 

In the present work, we instead tackle the more practical IoT condition of \emph{exogenous} traffic generation, in which a device may not always have fresh data to send. For this setting we provide the first analytical characterization of the average AoI of the Capetanakis scheme with gated access \cite{Capetanakis79}, resorting to a Markovian approach.  Our study captures the non-trivial and  fundamentally dynamic behavior of the protocol, where the number of users trying to access the channel is driven by the duration of the previous collision resolution time, and explores its implication on AoI. To this aim, we also consider a variation of the strategy, and study the idea of \emph{early termination}. Accordingly, a contention resolution phase can be truncated once a maximum number of slots is reached, even if not all the users originally involved in the collision have been decoded. The solution triggers a fundamental trade-off between reliability and latency, and allows us to shed light on the value of keeping retransmitting packets that progressively become stale, under the more age-challenging exogenous traffic. By providing the optimal maximum duration of contention resolution under any packet generation rate, the analysis offers some useful protocol design guidelines. 

\section{System Model}
\label{sec:sysModel}

We focus on a population of \nodes\ terminals (users), which share a common wireless channel to send time-stamped messages to a common receiver. Time is slotted, and the slot duration is set to accomodate the transmission of a packet. Channel access follows a random-access approach, and a collision channel model is considered. Accordingly, if a single packet is sent over a slot, it is successfully decoded, whereas if two or more nodes access the channel simultaneously, no data is retrieved. Terminals implement the Capetanakis-Tsybakov-Mikhailov algorithm with gated access (CTM) \cite{Capetanakis79}. In the interest of space, we provide in the following a simple description of the scheme, referring the reader to the vast available literature for further details, e.g., \cite{Massey81,Mathys85,Giannakis07}. 

The protocol works as sequence of contention resolution intervals (CRIs), composed by a variable number of slots. 
At the end of a CRI, all nodes that have a new packet to send, will transmit with probability $1$ in the next slot. Three cases are possible: i) no terminal has data and the slot remains \emph{idle}; ii) only one user has data (\emph{singleton}); iii) two or more terminals transmit, leading to a \emph{collision}. The outcome is broadcasted by the receiver to all nodes via an instantaneous and ideal ternary feedback. If an idle or singleton slot was experienced, the newly started CRI immediately comes to an end. If, instead, a collision took place, all nodes involved in it initiate a resolution procedure, wheras any other terminal remains silent until the CRI will be concluded. Each terminal that collided, flips a binary fair coin, choosing whether to (a) attempt transmission again, or (b) keep idle. In the latter case, the terminal will refrain from accessing the channel until \emph{all} the users who decided for option (a) successfully delivered their message. When this condition is met, the user will transmit in the next slot with probabiltiy $1$. Instead, any user in the original collision who decided to attempt again (option (a)), will re-transmit its message in the very next slot. This set of rules is  iterated, with transmitting nodes splitting in two groups after a collision, and with any user who experiences a singleton slot exiting the contention and refraining from access until the next CRI. The algorithm allows for a simple and elegant implementation, originally proposed by Gallager \cite{Massey81}, using two counters that track the evolution of the contention and that are updated based on the feedback after each slot. Specifically, at the start of a CRI, every node in the network sets a \emph{global} counter to $1$. Moreover, any time a node involved in the CRI decides to remain idle after a collision (option (b)), it sets a \emph{local} counter to $1$. Terminals update both their global and local counters at the end of each slot, based on feedback: in case of an idle or singelton slot counters are decreased by $1$; in case of a collision, counters are increased by $1$. It is easy to prove that, whenever the \emph{local} counter of a node reaches $0$, the user shall transmit again in the next slot, as all terminals that chose to transmit when its idling decision was made have been resolved. Moreover, when the \emph{global} counter reaches $0$, the CRI terminates, with all users having been resolved. The algorithm allows for a simple implementation, originally proposed by Gallager \cite{Massey81}. Specifically, at the start of a CRI, every node in the network sets a \emph{global} counter to $1$. Moreover, any time a node involved in the CRI decides to remain idle after a collision (option (b)), it sets a \emph{local} counter to $1$. Terminals update both their global and local counters at the end of each slot, based on feedback: in case of an idle or singelton slot counters are decreased by $1$; in case of a collision, counters are increased by $1$. It is easy to prove that, whenever the \emph{local} counter of a node reaches $0$, the user shall transmit again in the next slot, as all terminals that chose to transmit when its idling decision was made have been resolved. Moreover, when the \emph{global} counter reaches $0$, the CRI terminates, with all users having been resolved. 

In the remainder, we will refer to this classical scheme as the \emph{plain CTM} algorithm. We remark that the protocol has two inherent properties. On the one hand, all nodes that join a CRI are resolved, i.e., any packet that is sent is eventually delivered. On the other hand, this leads to potentially long resolution intervals, as the duration of a CRI is unbounded.\footnote{We note that the system does not undergo the well-studied instability issues of CTM \cite{Massey81}, as we deal with a finite population.} 
To explore this trade-off, we also study a variation of the scheme, denoted as \emph{CTM with early termination} (CTM-ET). The protocol works as the plain CTM, with the only difference that, if the CRI duration reaches a maximum length of \Lmax\ slots, the contetion is terminated even if not all nodes have yet been resolved. When this happens, users that did not succeed drop their packet, and do not retransmit it. The CRI is concluded, and the next one starts as before (with all counters being reset).

The system description is completed by considering how terminals generate traffic. In this respect, we assume that, at each slot, a user independently generates a new time-stamped message with probability \pGen, and places it in a one-packet sized buffer. The message is removed from the buffer at the start of the CRI over which it is transmitted, or if it replaced by the generation of a new packet. \footnote{In other words, we consider a buffering policy with pre-emption in waiting. The choice of a one-sized buffer is driven by the interest in age of information, as transmitting an older packet whenever a new one is available would only lead to delivery of staler information.} 
Accordingly, at the end of a CRI of duration $\ell$ slots, a node will have a message to send (and will thus contend in the next CRI) if it has generated at least once in the past $\ell$ slots, i.e., with probability 
\begin{align}
    \Gamma_\ell := 1 - (1-\pGen)^\ell.
    \label{eq:gamma}
\end{align}
The model is inspired by practical IoT networks, in which a transmitter is often fed with sensor readings and cannot control when these are produced. We are thus concerned with an \emph{exogenous} rather than the classical \emph{generate-at-will} traffic.

In this setting, we are interested in the ability of the access schemes to maintain an up-to-date perception at the receiver. We thus focus without loss of generality on a reference user, and consider the instantaneous AoI at the receiver \cite{Yates20_Survey}. This is defined at time $t$ as $\delta(t) = t - \tau(t)$, where $\tau(t)$ is the generation time of the last packet successfully received from the user. We then measure performance in terms of the average AoI: 
\begin{align}
    \Delta = \lim_{t\to\infty}\frac{1}{t}\int_0^t \delta(v) dv.
\end{align}

\emph{Notation:} We denote a discrete random variable (r.v.) and its realization by upper- and lower-case letters, respectively, e.g. $X$, $x$. The probability mass function of $X$ is indicated as $p_X(x)$, with straightforward extension to conditional distributions. Whenever clear from the context, we will omit the subscript for brevity. Moreover, we denote by calligraphic font the probability generating function (PGF) of $X$, i.e., %$\mathcal X(z) = \expOp\left[ z^X \right] = \sum\nolimits_{x} p(x) z^x$, 
\begin{align}
    \mathcal X(z) = \expOp\left[ z^X \right] = \sum\nolimits_{x} p(x) z^x
\end{align}
where the summation spans the whole alphabet of $X$. 
\section{Preliminaries}
\label{sec:prelim}

We start by stating some results that will be useful in the computation of the average AoI. 

\begin{lemma}
    For plain CTM, let $\tilde L$ be the r.v. denoting the duration in slots of a CRI, and let $\mathcal L_u(z)$ be its PGF conditioned on having $u$ devices contending at the start of the CRI. Then, the following recursion holds:
    \begin{align}
        \mathcal L_u(z) = \frac{z}{2(2^{u-1} -z^2)} \, \sum_{i=1}^{u-1} \binom{u}{i} \mathcal L_i (z) \mathcal L_{u-i}(z)  
        \label{eq:pgfL}
    \end{align}
    with $\mathcal L_0(z) = \mathcal L_1(z) = z$.
\end{lemma}
\begin{IEEEproof}
Following standard arguments, e.g., \cite{Mathys85}, the PGF for a binary fair split can be expressed as
\begin{align}
    \mathcal L_u(z) = \sum\nolimits_{i=0}^{u}\binom{u}{i} \frac{z}{2^u} \, \mathcal L_i(z)\, \mathcal L_{u-i}(z).
    \label{eq:pgfL_basic}
\end{align}
The statement follows by rearranging \eqref{eq:pgfL_basic}, and by observing that when a single or no user contend the CRI is terminated after one slot, so that $\sum\nolimits_{\tilde \ell=1}^{\infty} p(\tilde \ell)z^{\tilde \ell} = z$.
\end{IEEEproof}

\begin{remark}
    The recursion in \eqref{eq:pgfL} allows to compute the conditional PFG of $\tilde L$ for any value of $u$. Moreover, evaluating the PFG in \mbox{$z=e^{-j 2 \pi k/\Lmax}$}, for $k\in\{1,\dots,\Lmax\}$, provides the discrete Fourier transform (DFT) of the sequence $p_{\tilde L\givenS U}(\tilde \ell\givenS u)$ for $\tilde \ell \in \{1,\dots,\Lmax\}$. The PMF %of $\tilde L$ conditioned on $u$ 
    can thus simply be obtained by taking the inverse DFT (IDFT) of the sequence.
\end{remark}

\begin{lemma}
    For plain CTM, focus on a node contending over a CRI, and let $\tilde D$ denote the number of slots between the start of the CRI and the moment the node is decoded, conditioned on having $M{\in}\{0,\dots,\nodes{-}1\}$ other users contending. The PFG of $\tilde D$ conditioned on $M$ satisfies the recursion $\mathcal D_1(z)=z$, and 
    \begin{align}
        \mathcal D_{m+1}(z) &= \frac{z}{2^{m+1}-z(z+1)} \Big[ z(1+ \mathcal L_m(z)) \\&+ \sum\nolimits_{i=1}^{m-1} \binom{m}{i} \left(\mathcal D_{i+1}(z) + \mathcal L_i(z)\,\mathcal D_{m-i+1}(z)\right)\Big]
    \end{align}
\end{lemma}
\begin{IEEEproof}
    The conditional PGF of $\tilde D$ was originally derived in \cite{Giannakis07}. Computing it for the binary fair splitting case and rearranging the terms leads to the recursion.
\end{IEEEproof}
Also in this case, the PMF $p_{\tilde D\givenS M}(\tilde d\givenS m)$, can be computed efficiently by taking an IDFT of the sequence $\mathcal D_m(e^{-j2\pi k/\nodes})$ for $k\in\{0,\dots,\nodes{-}1\}$, as streamlined in Remark 1.

\subsection{CTM with Early Termination (CTM-ET)}
Let us now focus on the case of early termination, and let $L$ be the r.v. denoting the duration of a CRI. The PMF of $L$ conditioned on the number of contending users readily follows from the plain CMT case, as
\begin{align}
    p_{L\givenS U}(\ell\given u) =
    \begin{cases}
        p_{\tilde L\givenS U}(\tilde \ell \given u) & \quad \ell < \Lmax\\[.3em]
        \sum_{\tilde \ell=\Lmax}^\infty p_{\tilde L\givenS U}(\tilde \ell \given u) & \quad \ell = \Lmax
    \end{cases}
    \label{eq:pLGivenU}
\end{align}
The dynamic behavior of CCRA over subsequent CRIs can further be captured by means of a few conditional PMFs. The first characterizes the random number of users $U_i$ accessing the $i$-th CRI, conditioned on the duration $L_{i-1}$ of the preceding contention period. Recalling that each node transmits at the start of a CRI if it generated at least one packet over the previous one, we obtain the binomial distribution
\begin{align}
    p_{U\givenS L}(u_i\given \ell_{i-1}) = \binom{\nodes}{\,u_i\,} (\pGenCRI_{\ell_{i{-}1}})^{u_i} \left(1-\pGenCRI_{\ell_{i{-}1}}\right)^{\nodes{-}u_i}.
    \label{eq:pUGivenL}
\end{align}
Leaning on \eqref{eq:pUGivenL}, the duration of the $i$-th CRI conditioned on the previous contention follows as
\begin{align}
    p(\ell_i\given \ell_{i-1}) = \!\!\sum\nolimits_{u_i=0}^\nodes p_{L\givenS U}(\ell_i\given u_i) \,p_{U\givenS L}(u_i\given \ell_{i-1}).
    \label{eq:pLGivenL}
\end{align}
In other words, the stochastic process $L_n$ describing the duration of subsequent CRIs is Markovian, with one-step transition probabilities given by \eqref{eq:pLGivenL}. The stationary distribution, derived via standard methods, will be denoted by $\pi_\ell$, $\ell\in \{1,\dots,\Lmax\}$.

In the remainder, we often focus on a user of interest, and track the number of other nodes $M$ that access a CRI. In this case, the distribution conditioned on the previous CRI duration, $p_{M\givenS L}(m_i\given \ell_{i-1})$, is also binomial, i.e., $\text{Bin}(\nodes-1,\pGenCRI_{\ell_{i-1}})$. 

Finally, we introduce the following result:
\begin{lemma} Focus on a reference user contending over a CRI, and let $\varphi(m)$ be the joint probability that the CRI terminates after \Lmax\ slots without resolving all terminals, and the reference user is \emph{not} decoded. Then,
\begin{align}
    \varphi(m) = 1 - \sum\nolimits_{\tilde d=1}^{\Lmax} p_{\tilde D\givenS M}(\tilde d\given m).
        \label{eq:mu}
    \end{align}    
\end{lemma}
\begin{IEEEproof}
The result follows observing that the event corresponds to having the user decoded after slot \Lmax\ if plain CTM without early termination were employed. Leaning on the conditional PMF of $\tilde D$ obtained from Lemma 2, this can be expressed as $\sum_{d=\Lmax{+1}}^\infty p_{\tilde D\givenS M}(\tilde d\givenS m)$, or, equivalently \eqref{eq:mu}.
\end{IEEEproof}

\section{Average AoI Characterization}
\label{sec:analysis}
\begin{figure}
    \centering
    \includegraphics[width=.9\columnwidth]{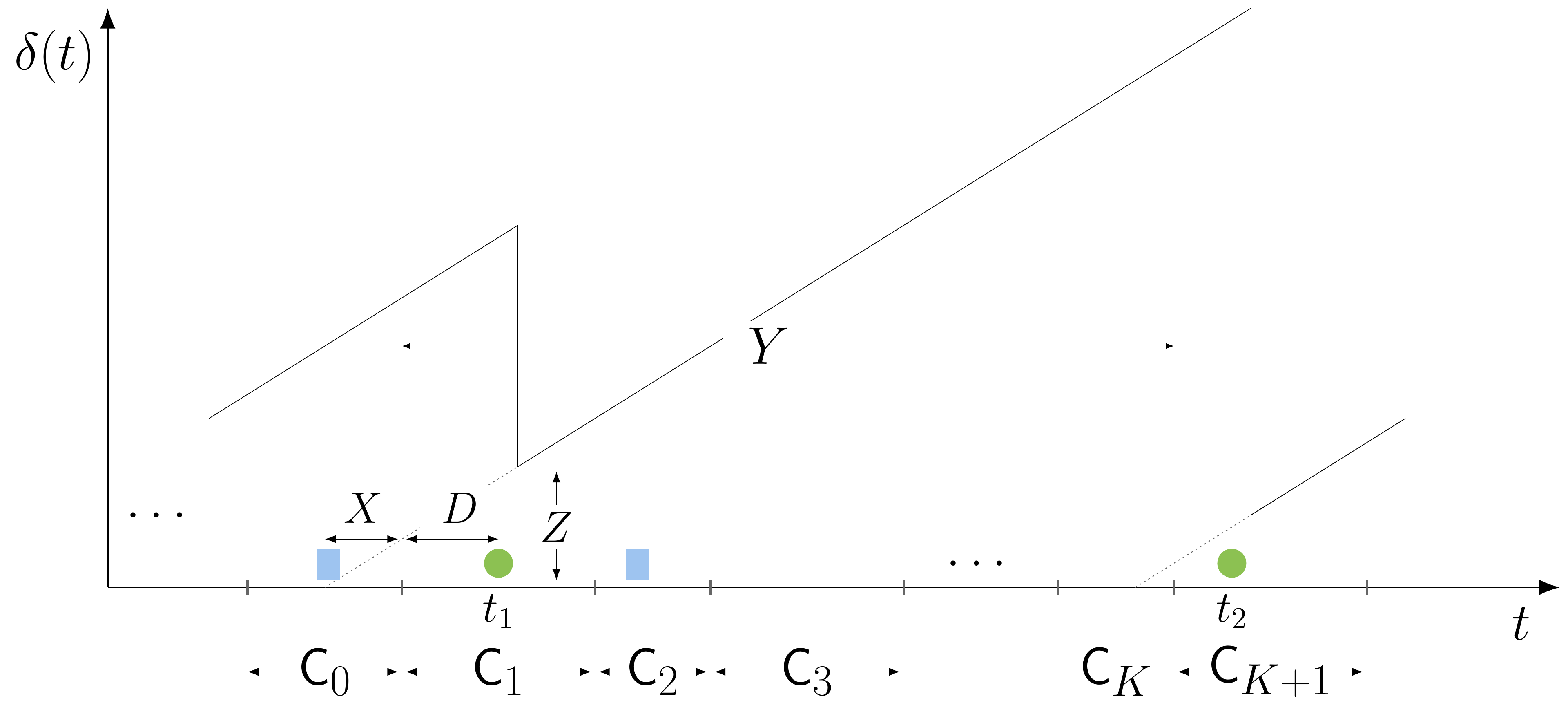}
    \caption{Time evolution example of AoI. Blue rectangles denote generation of a new packet at the user, whereas green dots a successful delivery. Due to early termination, the packet generated in CRI $\cri_2$ is not delivered over $\cri_3$.}
    \label{fig:timeline}
\end{figure}

We treat the case of early termination, clarifying in Remark 2 how the analysis can be adapted to plain CTM. Let us focus without loss of generality on a node of interest, and denote by $\Refresh_n$ the stochastic process describing its AoI inter-refresh time, i.e., the number of slots between two successive update deliveries by the device. Moreover, let $\Reset_n$ be the value at which the AoI is reset at the beginning of the $n$-th inter-refresh interval (see \figr\ref{fig:timeline}). Under the assumption that the processes are ergodic, we focus on their stationary behavior, and drop the subscript $n$ for readability. With this notation, following standard geometric arguments \cite{Munari23_TCOM,Yates20_Survey}, the average AoI of the node of interest can be expressed as
\begin{align}
    \Age = \frac{\expOp[\Reset \Refresh] + \expOp[\Refresh^2]/2}{\expOp[\Refresh]}.
    \label{eq:age_moments}
\end{align}
The calculation of \eqref{eq:age_moments} is non trivial. Indeed, the different CRIs within an inter-refresh period have durations that are not independent of each other, due to the dynamic behavior of CTM. Moreover, \Reset\ and \Refresh\ are also not independent, as the value of \Reset\ is influenced by the duration of the CRI which leads to the AoI reset, and this in turn impacts the first CRI in \Refresh. 

To tackle the problem, the timeline reported in \figr\ref{fig:timeline} shows the structure of an inter-refresh period. At the start, the reference node is idle, and generates a new update during CRI $\cri_0$. The blue rectangle shows the slot of the last message generated by the node during $\cri_0$, i.e., the time-stamp of the packet it will send over the subsequent CRI $\cri_1$. In turn, the green dot shows the slot in $\cri_1$ at which the device is decoded, resetting its AoI value to \Reset. This is followed by a number of CRIs during which the device does not deliver, until $\cri_K$ is reached, which sees the generation of a new update, sent and decoded over $\cri_{K+1}$. In this example, the inter-refresh time is $Y{=}\,t_2{-}t_1$. To ease calculations,  we introduce here a simplification, and express %$Y \simeq \sum_{i=1}^K \LCRI_i$
\begin{align}
    Y \simeq \sum_{i=1}^K \LCRI_i
\end{align}
where the r.v. $\LCRI_i$ denotes the number of slots composing CRI $\cri_i$.
The approach is motivated by noting that the formulation of $Y$ disregards the time between the end of $\cri_K$ and $t_2$, yet compensates this by adding the same statistical quantity for the previous refresh, i.e., the time between the start of $\cri_1$ and $t_1$. The tightness of the assumption will be verified in \secr\ref{sec:results}.

Before delving into the calculation of \eqref{eq:age_moments}, it is useful to derive the joint distribution of the duration of CRIs $\cri_0$ and $\cri_1$, i.e., $p(\lCRI_0,\lCRI_1)$. To this aim, consider the conditions leading to the 
start of an inter-refresh period: i) a CRI of duration $\ell_i$ slots, with at least one packet generated by the node, followed by ii) a CRI of duration $\ell_{i{+}1}$ in which the node successfully delivers. We denote the event probability as $\theta(\ell_i,\ell_{i{+}1})$, and have, for $\ell_{i{+}1}<\Lmax$
\begin{align}   
   \!\!\!\!\!\!\theta(\ell_i,\ell_{i{+}1}) \!=\! \pi_{\ell_i} \pGenCRI_{\ell_i} \sum\nolimits_{m=0}^{\nodes-1} p_{M\givenS L}(m\givenS \ell_i) p_{L\givenS U}(\ell_{i+1}\givenS m{+}1).
  \label{eq:jointC0C1}
\end{align}
In \eqref{eq:jointC0C1}, the factor $\pi_{\ell_i} \pGenCRI_{\ell_i}$ captures the stationary probability of event i), whereas the rest accounts for the probability of ii), i.e., having a CRI of $\ell_{i{+}1}$ slots, given $\ell_i$, and decoding the user of interest. The latter is obtained in turn by conditioning on the number $m$ of nodes other than the user that become active over the first CRI, and noting that for $\ell_{i{+}1}{<}\Lmax$, all users are decoded. When the maximum CRI duration is reached, instead, we have to account for two possibilities: all users are resolved in exactly $\Lmax$ slots, or the CRI is terminated, but the user of interest is decoded. Leaning on Lemma 3, we can then write 
\begin{align}   
    \theta(\ell_i,\Lmax) \!=\! \pi_{\ell_i}  \pGenCRI_{\ell_i} \!\sum_{m=0}^{\nodes{-}1} \!p_{M\givenS L}(m\givenS \ell_i) \big[p_{L\givenS U}(\Lmax\givenS m{+}1)-\varphi(m)\big].\\[-2em]
    \label{eq:jointC0C1_LMax}
 \end{align}
Based on \eqref{eq:jointC0C1}-\eqref{eq:jointC0C1_LMax}, we finally have the overall distribution
\begin{align}
    p(\lCRI_0,\lCRI_1) = \frac{\theta(\lCRI_0,\,\lCRI_1)}{\sum_{\ell_i,\ell_{i{+1}}}\theta(\ell_i,\ell_{i{+}1})}
    \label{eq:jointPMFL0L1}
\end{align}
where the summation is taken for $\ell_i,\ell_{i{+}1}\in\{1,\dots,\Lmax\}$.

\subsection{Statistical moments of the inter-refresh time}
\label{sec:irt}
Let us tackle the expected value of $Y$, conditioned on the duration of the first CRI $\cri_1$, denoted as $Y^1_{\lCRI_1} := \expOp[Y\given \LCRI_1{=}\lCRI_1]$. Leaning on the Markovian evolution of the contention durations, a first step analysis \cite{Gallager:StochasticProc} can be applied, to obtain
\begin{align}
    \begin{split}
    \!\!\!\!\!\!Y^1_{\lCRI_1} =& \,\,\lCRI_1 \!+\! \sum_{\lCRI=1}^{\Lmax} Y^1_{\lCRI} \cdot(1{-}\pGenCRI_{\lCRI_1}) \sum\limits_{m=0}^{\nodes{-}1} \!p_{M\givenS L}(m\givenS \lCRI_1) \,p_{L\givenS U}(\lCRI\givenS m) \\[-.2em]
    &\quad + Y^1_{\Lmax} \pGenCRI_{\lCRI_1} \sum_{m=0}^{\nodes-1} p_{M\givenS L}(m\givenS \lCRI_1) \, \varphi(m) .
    \end{split}
    \label{eq:firstStepY}
\end{align}
In \eqref{eq:firstStepY}, the overall duration is expressed as the sum of the first CRI ($\lCRI_1$ slots), and of the expected duration of an inter-refresh period starting with a CRI of length $\lCRI$, weighted by the probability for this to be undergone after $\mathsf C_1$. Consider the first row of \eqref{eq:firstStepY}. In this case, the reference user does not generate a packet over $\mathsf C_1$ (probability $1{-}\pGenCRI_{\lCRI_1}$), and the inter-refresh period will certainly continue. Following the usual approach, the probability for the next CRI to last $\lCRI$ can directly be obtained by conditioning on the number of nodes that will be active among the remaining $\nodes{-}1$. If, instead, the reference node does generate at least one packet during $\mathsf C_1$ (second row of the expression), the inter-refresh period will continue only if the user will not successfully deliver the message over the next CRI. This requires the CRI to be of duration \Lmax\ slots, to be terminated, and for the user not to be decoded. Conditioning on the number of other contenders, this probability is exactly captured by $\varphi(m)$ in Lemma 3.
Overall, \eqref{eq:firstStepY} can be computed for any starting value of $\lCRI_1$, providing a full-rank system of $\Lmax$ linear equations in \Lmax\ unknowns, whose solution can be derived with standard methods and provides the sought first order moment of $\Refresh$ conditioned on the length of $\cri_1$. 

The same reasoning can be applied to derive the second order moment of $Y$ conditioned on $\LCRI_1$, denoted as $Y^2_{\lCRI_1}$. In this case, the first-step approach leads to the system of equations
\begin{align}
    Y^2_{\lCRI_1} \!=& (\lCRI_1)^2 \!+\!\! \sum_{\lCRI=1}^{\Lmax}\! \alpha(\lCRI_1,\lCRI) (1{-}\pGenCRI_{\lCRI_1}) \sum\limits_{m=0}^{\nodes{-}1} \!p_{M\givenS L}(m\givenS \lCRI_1) p_{L\givenS U}(\lCRI\givenS m) \\[-1.5em]
    &\quad + \alpha(\lCRI_1,\Lmax) \,\pGenCRI_{\lCRI_1} \sum_{m=0}^{\nodes-1} p_{M\givenS L}(m\givenS \lCRI_1) \, \varphi(m) 
    \label{eq:firstStepY2}
\end{align}
where we have introduced for compactness the ancillary quantity $\alpha(\lCRI_1,\lCRI) {:=} (2\lCRI_1 Y^1_{\lCRI} + Y^2_{\lCRI})$.

\subsection{AoI reset value}
Consider now $\expOp[ZY]$, conveniently expressed as
\begin{align}
    \!\!\!\!\!\expOp[ZY] = \sum\nolimits_{\lCRI_0,\lCRI_1} \expOp[ZY \given \LCRI_0=\lCRI_0,\LCRI_1=\lCRI_1] \, p(\lCRI_0,\lCRI_1).
    \label{eq:expZY}
\end{align}
Conditioned on the duration of the two CRIs, the age-reset value $Z$ and the inter-refresh time $Y$ become in fact independent. This comes from the fact that the former refers to a packet generated in $\cri_0$ and  decoded in $\cri_1$, whereas the latter is driven by the generation and delivery of subsequent packets, which is an independent process once the duration $\LCRI_1$ is fixed. The overall computation thus simplifies into the product of the two conditional expectations of $Y$ and $Z$. 

We focus first on the age-reset value, observing from \figr\ref{fig:timeline} that it can be expressed as $Z = X + D$. Here, $X$ is the number of slots elapsed between the packet generation and the end of $\cri_0$, and $D$ is the time between the start of $\cri_1$ and the delivery of the message.
To derive  the statistics of $X$, we note that the variable by definition does not depend on $\LCRI_1$. In turn, the PMF conditioned on $\LCRI_0$ can be written for any $x\in\{1,\dots,\lCRI_0\}$ as
\begin{align}
    p_{X\givenS \LCRI_0}(x\given \lCRI_0) = \left[\pGen(1-\pGen)^{x{-}1}\right]/\pGenCRI_{\lCRI_0}
    \label{eq:pXGivenL0}
\end{align}
where the numerator captures the probability of generating the message and then not producing updates for the remaining $x{-}1$ slots of $\cri_0$, whereas the denominator normalizes the distribution to having generated at least one packet.

\begin{figure}
    \centering
    \subfloat[]{
    \includegraphics[width=.47\columnwidth]{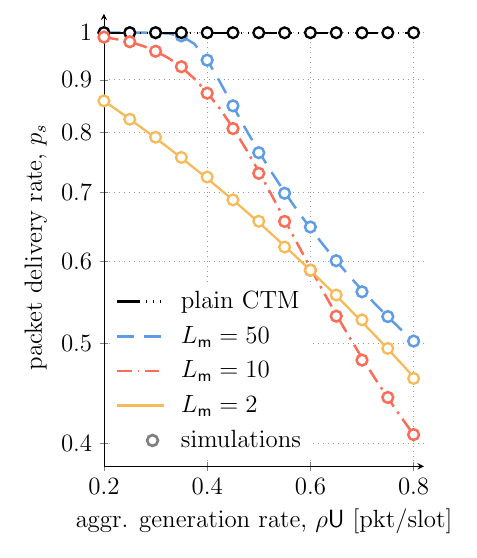}}
    \hspace{.5em}
    \subfloat[]{
    \includegraphics[width=.47\columnwidth]{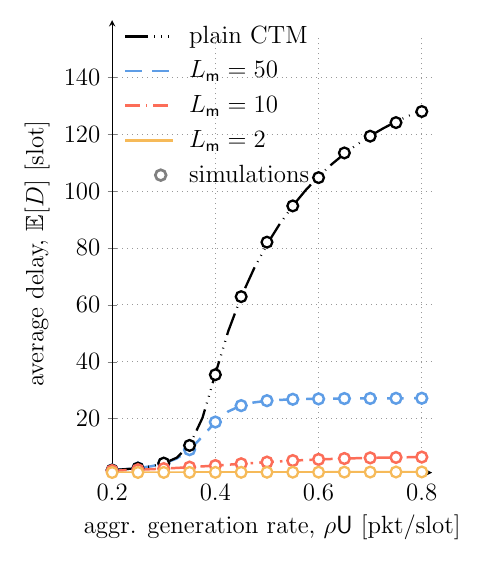}}
    \caption{(a) packet delivery rate, $p_s$ and (b) average delay, $\mathbb E[D]$, plotted against the aggregate generation rate $\pGen\nodes$.}
    \label{fig:psAndDel}
\end{figure}

On the other hand, the conditional distribution of $D$ given $\LCRI_0$ and $\LCRI_1$ is non-trivial. In particular, determining the decoding slot of a message conditioned on the duration of the CRI it is sent over introduces inter-dependencies that prevent a direct application of the standard recursions employed for tree-based algorithms. In view of this, we resort to an approximation, and consider the term $\expOp[D\givenS \LCRI_0=\lCRI_0,\LCRI_1=\lCRI_1] \simeq \expOp[D\givenS\LCRI_0=\lCRI_0]$. 
The approach stems by observing that conditioning on $\LCRI_0$ already determines the number of contenders over $\cri_1$, and thus drives its duration. The approximation proves to be extremely tight for all configurations of arrival rate and \Lmax, as highlighted in \figr\ref{fig:psAndDel}(b). Following this assumption, we work with the conditional distribution $p_{D\givenS \LCRI_0}(d\givenS \lCRI_0)$, which, in turn, can be obtained from $p_{\tilde D\givenS M}(\tilde d\givenS m)$ (Lemma 2) as
\begin{align}
    \!\!\!\!\!\!p_{D\givenS \LCRI_0}(d\given \lCRI_0) = \frac{\sum_{m=0}^{\nodes{-}1} p_{\tilde D \givenS M}(\tilde d \givenS m) \, p_{M\givenS L}(m\givenS \lCRI_0)}{\sum_{i=1}^{\Lmax}\sum_{m=0}^{\nodes{-}1}p_{\tilde D \givenS M}(\tilde i \givenS m) \, p_{M\givenS L}(m\givenS \lCRI_0)}.
    \label{eq:pDGivenL0}
\end{align}
For $d{\in}\{1,\dots,\Lmax\}$, the numerator in \eqref{eq:pDGivenL0} reflects the condition on $\LCRI_0$ in terms of the number of contenders $M$ over $\cri_1$, while the denominator provides the normalization for considering CRIs (and thus decoding) of maximum $\Lmax$ slots.

The distributions in \eqref{eq:pXGivenL0} and \eqref{eq:pDGivenL0} allow to compute the (approximated) conditional expectation $\expOp[Z\givenS \LCRI_0{=}\lCRI_0,\LCRI_1{=}\lCRI_1]$. On the other hand, $Y$ only depends on $\LCRI_1$, whose conditional expectation has been captured in \secr\ref{sec:irt}. Plugging the product of the two into \eqref{eq:expZY} and leveraging the joint PMF $p(\lCRI_0,\lCRI_1)$ in \eqref{eq:jointPMFL0L1} finally provides the sought first order moment, concluding the calculation of the average AoI in \eqref{eq:age_moments}.

\begin{remark}
    The analysis presented so far for early-termination can be extended straightforwardly to plain CTM. In this case, besides considering summations that span $\ell{\in}\mathbb N$ and referring to the r.v.s $\tilde L$ and $\tilde D$, an active user is always decoded within a CRI. This simplifies the analysis, as, for instance, the final summation terms in \eqref{eq:firstStepY} and \eqref{eq:firstStepY2} can be omitted, describing early-terminated CRIs in which an active node does not deliver. For all practical purposes, however, the plain CTM performance can be obtained by implementing the reported calculations for a sufficiently large value of \Lmax.
\end{remark}

\section{Results and Discussion}
\label{sec:results}

To evaluate performance, we focus on a setting with $\nodes{=}100$ users. Configurations with different \nodes\ were also studied, leading to similar trends, and are not reported for brevity. In the following plots, lines refer to the analysis of Sec.~\ref{sec:analysis}, whereas circle markers show the outcome of Montecarlo simulations, implementing the CTM protocol in all its details. 

To gather preliminary insights on the trade-offs induced by early termination, we report in \figr\ref{fig:psAndDel}(a) the packet delivery rate $p_s$, i.e., the probability that a node initiating a packet transmission succeeds within \Lmax, and in \figr\ref{fig:psAndDel}(b) the average delay undergone by a delivered message, i.e., $\expOp[D]$. Both quantities are shown against the aggregate packet generation rate, $\pGen\nodes$. 
The black, dash-dotted line denotes plain CTM, whereas the colored lines refer to the CTM-ET, for different \Lmax. Consider first the packet delivery rate. The metric can be derived analytically as
\begin{align}
    p_s = \big[\sum\nolimits_{\ell_i,\ell_{i{+}1}}\theta(\ell_i,\ell_{i{+}1}) \big]/\sum\nolimits_{\ell_i} \pi_{\ell_i} \pGenCRI_{\ell_i}.
    \label{eq:ps}
\end{align}
In \eqref{eq:ps}, the numerator captures the probability for a node to generate a packet over the $i$-th CRI, and to deliver it over the next one, as computed in \eqref{eq:jointC0C1}-\eqref{eq:jointC0C1_LMax}, whereas the denominator normalizes to having the node transmit over the $(i{+}1)$-th CRI. By construction, all packets are delivered with the plain CTM ($p_s{=}1$). Instead, early termination may lead to losses, and the effect is more pronounced as \Lmax\ reduces. Eventually, for $\Lmax{=}2$, the protocol behaves like slotted ALOHA, and $p_s$ approaches for $\pGen\nodes=0.8$ the value $e^{-\pGen\nodes} = 0.45$. More interestingly, the configuration $\Lmax{=}10$ experiences an even higher loss rate for large \pGen, in spite of allowing more time for collision resolution. This effect is rooted in the dynamic behavior of CTM across CRIs. An intuition is obtained recalling that, at the end of a CRI of $\ell$ slots, on average $\nodes(1{-}(1{-}\pGen)^\ell)$ users will have updates to send in the next slot. Assume that CRIs run until they are terminated. When $\Lmax{=}2$, on average $\sim 1.6$ nodes will contend, whereas for $\Lmax{=}10$ this increases to $\sim 7.8$. As a result, in the latter case, a much larger collision set has to be resolved, and it is likely that not all the user will be retrieved within the available $10$ slots, leading to a lower delivery rate. 
The increased losses triggered by an early termination, however, also reduce the expected delay of delivered messages, as shown by \figr\ref{fig:psAndDel}(b).\footnote{Analytical results were in this case obtained by removing the conditioning on $\LCRI_0$ in \eqref{eq:pDGivenL0} via the marginalization of \eqref{eq:jointPMFL0L1}, and by taking the expectation.} In this respect, the excellent match between simulations and analysis confirms the validity of the approximation introduced in \secr\ref{sec:analysis}. 

\begin{figure}
    \centering
    \includegraphics[width=.93\columnwidth]{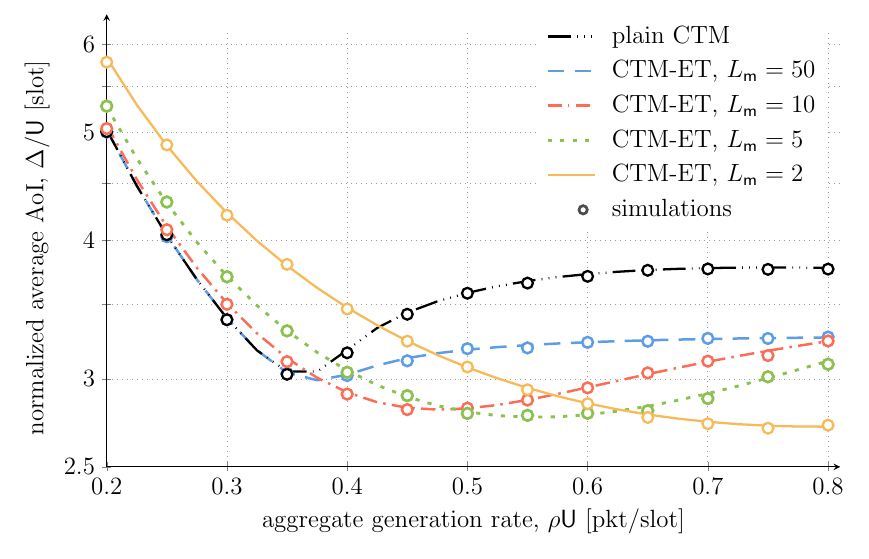}
    \caption{Normalized average AoI, $\Age/\nodes$ vs aggregate generation rate, $\pGen\nodes$.}
    \label{fig:aoiVsLambda}
\end{figure}

The fundamental trade-off between latency and update delivery eventually drives the AoI performance. The behavior of the metric, normalized to the number of nodes, is reported in \figr\ref{fig:aoiVsLambda} as a function of $\pGen\nodes$. For plain CTM and for low arrival rates, large values of \Age\ are experienced, as nodes sporadically have new updates to send, leaving the receiver with stale information. In turn, high packet generation rates lead the system to operate over longer CRIs, where more nodes contend. Congestion leads to larger latencies, and thus to an increase of AoI. Notably, the minimum \Age\ is attained by plain CTM for an aggregate generation rate ${\sim} 0.347$, corresponding to peak throughput conditions \cite{Massey81}. When early termination is implemented, two effects can be noticed. On the one hand, a worse performance is attained for low \pGen, as AoI deteriorates of more than $15\%$ for $\Lmax{=}2$ compared to plain CTM. A complete collision resolution is thus especially beneficial in such conditions, allowing all nodes to deliver their message and avoiding potentially long periods of time without further updates generation. On the other hand, a truncation becomes convenient for larger values of $\pGen$. As highlighted by the plot, an AoI reduction of roughly $30\%$ is gained for $\Lmax{=}2$ compared to resolving all nodes. The result can be understood observing how plain CTM spends more time to resolve the larger initial collision sets experienced for high generation rates, leading to delivery of messages that have in the meantime become stale. In this respect, CTM-ET would drop packets whose time stamp is growing old, allowing nodes to attempt transmission of fresher updates that may have been generated in the meantime. 

\begin{figure}
    \centering
    \subfloat[]{
    \includegraphics[width=.47\columnwidth]{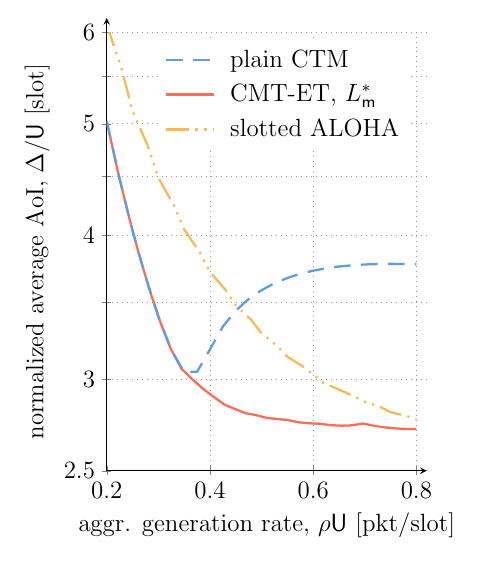}}
    \hspace{.5em}
    \subfloat[]{
    \includegraphics[width=.47\columnwidth]{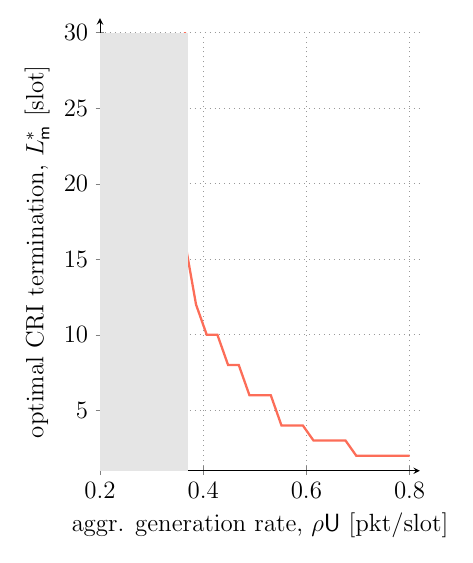}}
    \caption{(a) minimum normalized average AoI for the optimized CTM-ET, and (b) corresponding optimal \Lmax\ value.}
    \label{fig:optAge}
\end{figure}

The presented result provide thus a key take-away, pinpointing the value of a thorough collision resolution under sporadic generation, and prompting the value of a more aggressive early termination in harsher traffic conditions. This, in turn, triggers the natural question of how \Lmax\ shall be tuned in order to minimize AoI. We tackle this aspect in \figr\ref{fig:optAge}(a), which shows, for any $\pGen\nodes$, the minimum AoI obtained by optimizing over \Lmax\ (solid line). For reference, the performance of the plain CTM (dashed line) as well as of a standard slotted ALOHA with no contention resolution (dash-dotted line) are also reported. The latter is obtained via the well-known formulation $\Delta = 1/2 + [\pGen(1-\pGen)^{\nodes{-}1}]^{-1}$ \cite{Yates17:AoI_SA,Munari21_TCOM_AoI}. The improvement triggered by a proper application of the CTM-ET approach is evident, with significant reductions in AoI with respect to both benchmarks especially for intermediate values of \pGen. The corresponding optimal choices of \Lmax\ are shown in \figr\ref{fig:optAge}(b), where the gray-shaded region to the left corresponds to having $\Lmax{=\infty}$ (plain CTM). From this standpoint, the presented framework provides a useful tool for protocol parameter tuning.

\section{Conclusions}
\label{sec:conclusions}

We presented an analytical characterization of the average AoI under the Capetanakis tree-based random access algorithm with gated access, in the presence of exogenous traffic. Via a Markovian approach, we capture the coupling between contention dynamics and information freshness. Moreover, we proposed a truncated resolution mechanism, revealing a trade-off between latency and reliability that becomes critical under age-sensitive traffic. The resulting insights provide design guidelines for IoT systems employing feedback-based collision resolution in scenarios where timely updates are crucial.

\bibliographystyle{IEEEtran}
\bibliography{IEEEabrv,biblio_RandomAccess,biblio_AoI}

\end{document}